\documentclass[preprintnumbers,amsmath,amssymb]{revtex4}

\usepackage{graphicx}

\begin{document}


\title{TYPICAL BEHAVIOUR OF THE GLOBAL ENTANGLEMENT OF
AN OPEN MULTIQUBIT SYSTEM IN A NON-MARKOVIAN REGIMEN}

\author{A.P. Majtey}
\address{Instituto Carlos I de F\'{\i}sica Te\'orica y
Computacional and Departamento de F\'{\i}sica At\'omica Molecular
y Nuclear, Universidad de Granada, Granada, E-18071, Spain.}
\author{A.R. Plastino}
\address{Instituto Carlos I de F\'{\i}sica Te\'orica y
Computacional and Departamento de F\'{\i}sica At\'omica Molecular
y Nuclear, Universidad de Granada, Granada, E-18071, Spain. \\
Universidad Nacional de La Plata, CREG-UNLP, C.C. 727, 1900 La
Plata, Argentina}

\begin{abstract}
We investigate the decay of the global entanglement, due
to decoherence, of multiqubit systems interacting with
a reservoir in a non Markovian regime. We assume that
during the decoherence process each qubit of the system
interacts with its own, independent environment.
Most previous works on this problem focused on particular
initial states or families of initial states amenable
of analytical treatment. Here we determine numerically
the typical, average behaviour of the system corresponding
to random initial pure states uniformly distributed (in the
whole Hilbert space of $n$-qubit pure states) according
to the Haar measure. We study systems consisting of
3, 4, 5, and 6 qubits. In each case we consider also the
entanglement dynamics corresponding to important particular
initial states, such as the GHZ states or multiqubit states
maximizing the global entanglement, and determine in which
cases any of these states is representative of the average
behaviour associated with general initial states.

\vskip 1cm

\noindent
keywords: decoherence, non Markovian dynamics, entanglement

\vskip 0.5cm

\noindent
pacs:  03.65.Yz; 03.67.Mn; 03.67.-a

%
%

\end{abstract}

\maketitle

\section{Introduction}

Decoherence is a quantum phenomenon that plays an important role
in connection both with the foundations of quantum physics and
with its technological applications \cite{S05,NC00,BP02}.
Decoherence is inextricably linked to another basic ingredient
of the quantum world: quantum entanglement \cite{BZ06,BCS04,BCS07,TMB10}.
Indeed, the diverse effects associated with
the phenomenon of decoherence are due to the development of
entanglement between the system under consideration  and its
environment. This entanglement, arising from the interaction
between an imperfectly isolated system and its surroundings,
leads to the gradual disappearance of various quantum features
exhibited by the system. These effects are at the core of
the nowadays orthodox, decoherence-based explanation of the
quantum-to-classical transition \cite{S05}.

One of the most important consequences of decoherence is that the
(internal) entanglement between the different parts of a composite
system undergoing decoherence tends to decay. This manifestation
of decoherence constitutes one of the main hurdles that has to be
overcome for the implementation of quantum technologies that
require entangled states as a resource \cite{NC00}. The decay of
entanglement under decoherence has been the focus of an intense
research activity in recent years
\cite{SK02,CMB04,DB04,HDB05,ACCAD08,AlmeidaEtAl07,SallesEtAl08,AolitaEtAl09,GBB08,BGB08,CTP09,BMPCP08,Siomau12}.
A particularly interesting feature of this process, first pointed
out by Zyczkowski and the Horodeckis in a pioneering effort almost
ten years ago \cite{ZHHH01}, is that sometimes entanglement
vanishes completely in a finite time. This remarkable phenomenon,
known as entanglement sudden death (ESD), has been the subject of
several theoretical and experimental studies
\cite{YE04,YE09,QJ08,Weinstein09,LRLSR08,BLFC07,BLFC08,ZMZXG10,MXA10,HPP09,HIPZ,Cole10,ZWZG09,MMPS10,HZSK10,TALO10}.

Studies of ESD in Markovian scenarios suggested that, once
entanglement has vanished due to decoherence, this valuable
resource is irremediably lost. However, the decoherence process
in non-Markovian regimes sometimes gives rise to an interesting
new effect: entanglement sudden revival.
The sudden revival of entanglement after a finite time interval
with zero entanglement following an ESD event has been recently
investigated in \cite{BLFC07,BLFC08}. These works considered a
system consisting of two independent qubits interacting with an
environment at zero temperature in a non-Markovian regime. The
entanglement dynamics of open multiqubit systems in non-Markovian
regimes has also been considered in various recent works
\cite{ZMZXG10,MXA10,HIPZ}. In particular, multiqubit systems
evolving from initial $GHZ$ or $W$ states have been studied in
detail in \cite{ZMZXG10,MXA10}. On the other hand, the average,
typical entanglement-related features corresponding to some
families of initial states of two-qubit, three-qubit, and
four-qubit systems have been investigated in \cite{HIPZ}.

The phenomenon of entanglement sudden revival allowed
by non-Markovian regimes deserves detailed scrutiny: it
is of clear interest from the fundamental point of
view and may also have technological implications.
 The aim of the present contribution is to explore
(statistically) the typical behaviour of the entanglement
exhibited by multipartite
systems interacting with the above mentioned environment.
In contrast to previous studies of entanglement evolution
in non-Markovian regimes, that focused on particular states
or particular families of states, we will explore the average,
typical entanglement dynamics associated with random, general
initial pure states uniformly distributed according to the Haar
measure. We are also going to consider the average behaviour
corresponding to mixed initial states consisting of a mixture
of a (random) pure state and the totally mixed state.

The article is organized as follows. A brief description
of the system studied here is provided in Section II.
Our main findings concerning the typical entanglement
evolution of $n$-qubit systems are detailed in
Section III. Finally, some conclusions are drawn in
Section IV.

\section{System-Reservoir Model}

We are going to explore the typical, average features
characterizing the entanglement dynamics of an open
system consisting of $n$ independent qubits evolving
in a non-Markovian regime. These
$n$-qubits evolve from an initial entangled state
generated through some previous, unspecified,
interacting process. Each of the above mentioned
qubits experiences the same decoherence process,
which is due to the interaction of the qubit with
its own environment. The evolution of the density matrix
$\rho_i$ of the $i$-qubit is given by a completely positive
trace-preserving map that admits a Kraus representation
of the form,

\begin{equation} \label{dinunqu}
\rho_i(0) \longrightarrow
 \rho_i(t) = \sum_{j=1}^{M} E_{j \, i} \
 \rho_i (0) \  E_{j \, i}^\dag,
\end{equation}

\noindent where $E_j\,\,\,j=1,\ldots, M$ constitute
an appropriate set of Kraus operators yielding a
complete description of the single qubit's dynamics.
The evolution of the total density matrix $\rho$
corresponding to the entire $n$-qubit system can
also be expressed in terms of the above Kraus
operators,

\begin{equation}\label{kraus-evolution}
\rho(0) \longrightarrow
 \rho(t) \, = \,  \sum_{i...j} E_{i \, 1} \otimes \ldots
 \otimes E_{j \, n} \ \rho(0) \  [ E_{i \, 1} \otimes \ldots \otimes E_{j \, n}]^{\dag}.
\end{equation}

We are going to assume that the dynamics of each qubit
of the system is determined by a ``qubit + reservoir''
interaction given by the paradigmatic Hamiltonian
\cite{BLFC07},

\begin{equation} \label{Hamiltonian}
H=\omega_0\sigma_+\sigma_-+\sum_k\omega_kb_k^{\dag}b_k
+\sum_kg_k(\sigma_+b_k+\sigma_-b_k^{\dag}),
\end{equation}

\noindent
where $\omega_0$ is the transition frequency of an atomic two level
system (qubit), $\sigma_+$, $\sigma_-$ are the atomic raising and
lowering operators, $b_k^{\dag}$, $b_k$ are the creation and
annihilation operators of the reservoir's  (electromagnetic)
$k$-mode, and $\omega_k$, $g_k$ denote the corresponding frequency
and coupling constant (with the qubit). A possible physical
realization of a system governed by the Hamiltonian (\ref{Hamiltonian})
is given by a two-level atom interacting with a reservoir consisting
of the electromagnetic modes of a high-$Q$ cavity. Following
\cite{BLFC07} we adopt for the reservoir an effective spectral
density of the form

\begin{equation}
J(\omega)=\frac{1}{2\pi}
\frac{\gamma_0\lambda^2}{(\omega_0-\omega)^2+\lambda^2}.
\end {equation}

\noindent The parameter $\lambda$, defining the spectral width of
the coupling, is then connected to the reservoir correlation time
$\tau_b$ trough $\tau_b\sim\lambda^{-1}$. On the other hand, the
parameter $\gamma_0$ is associated with the system's relaxation
time $\tau_R$ according to the relation $\tau_R\sim\gamma_0^{-1}$
(see \cite{BLFC07} for details). Non-Markovian dynamics is
obtained in the strong-coupling regime given by the inequality
$\gamma_0 > \lambda/2$. The time dependent density matrix
describing the behaviour of a single qubit has the form

\begin{equation} \label{single-qb-evolution}
\rho(t)= \Bigl( \rho_{00}(0)+\rho_{11}(0)(1-p_t) \Bigr)
 |0\rangle \langle 0| \, + \,
  \rho_{01}(0)\sqrt{p_t}  |0\rangle \langle 1| \, + \,
   \rho_{10}(0)\sqrt{p_t} |1\rangle \langle 0| \, + \,
    \rho_{11}(0)p_t |1\rangle \langle 1|,
\end{equation}

\noindent
where the $\rho_{ij}(0)$ denote the elements of the
initial density matrix, and $p_t$ is given by

\begin{equation} \label{rotunqubit}
p_t=e^{-\lambda t}\left[\cos\left(\frac{d
t}{2}\right)+\frac{\lambda}{d}\sin\left(\frac{d
t}{2}\right)\right]^2,
\end{equation}

\noindent
where $d=\sqrt{2\gamma_0\lambda-\lambda^2}$ \cite{BLFC07}.
In the present work we are going to focus on the non-Markovian
regime, setting $\lambda=0.01\gamma_0$ in all our computations.

It can be verified, after some algebra, that the evolution
of the single qubit reduced density matrix elements given
by (\ref{single-qb-evolution}) admits a representation of
the form (\ref{dinunqu}) involving the following pair of
Kraus operators,

\begin{eqnarray}\label{NM-kraus}
E_0 \, &=& \, |0\rangle \langle 0| \, +\, \sqrt{p_t} |1\rangle \langle 1| \cr
E_1 \, &=& \, \sqrt{1-p_t}  |0\rangle \langle 1|,
\end{eqnarray}

\noindent
By recourse to equation (\ref{kraus-evolution})
the alluded Kraus representation of the single-qubit
dynamics can be used to obtain the full dynamics
characterizing the whole $n$-qubit system.

\section{Typical Entanglement Dynamics for $n$ Qubits}

\subsection{Preliminaries}

We are going to study the average entanglement dynamics associated with
random initial multiqubit pure states. These initial states are
of the form

\begin{equation}
|\Psi\rangle \, = \, U^{(n)} |\Psi_0 \rangle,
\end{equation}

\noindent
where $|\Psi_0 \rangle$ is a fixed $n$-qubit state, and $U^{(n)}$
is a random unitary transformation acting on the multiqubit
system. The unitary transformations $U^{(n)}$ are generated
according to the uniform distribution determined by the
Haar measure (see \cite{BCPP02} and references therein).
To evaluate the average properties of the entanglement
dynamics of the $n$-qubit system we determine the entanglement
evolutions corresponding to many initial states generated
according to the procedure explained, and compute the
concomitant mean entanglement evolution.

We also investigate the entanglement dynamics of random
mixed states of the form,

\begin{equation}\label{mixed-states}
\rho_{mixed}= x|\Psi\rangle\langle\Psi|+\frac{(1-x)}{2^n}
\, {\mathbb I},
\end{equation}

\noindent
with $0 \le x \le 1$. Here $x$ is the purity
of the multipartite state, ${\mathbb I}$ is
the $n\times n$ identity operator, and $|\Psi\rangle$ is a random
pure state generated following the same procedure as the one
used to generate random initial pure states. \\

To quantify the entanglement of a multipartite state one can deal
with measures defined as the average of bipartite entanglement
measures over all the possible bi-partitions of the system. This
is usually known as global entanglement and a considerable amount
of research has recently been devoted to its study (see
\cite{BSSB05,BPBZCP07} and references therein). The global
entanglement $E$ can be expressed as follow,

\begin{eqnarray}\label{Ent-measure}
E &=& \frac{1}{[n/2]} \sum_{m=1}^{[n/2]} E^{(m)}, \\
E^{(m)} &=& \frac{1}{K_{bipart}^{(m)}} \, \sum_{bipart} E(\rm
bipart.). \label{Entsub}
\end{eqnarray}

\noindent In equations (\ref{Ent-measure}-\ref{Entsub}) the
possible bi-partitions of the $n$-qubit system are grouped in
families corresponding to partitions into two sub-systems with
given numbers of constituting qubits. $E^{(m)}$ denotes the
average entanglement associated with the particular family
comprising the $K_{bipart}^{(m)}$ non-equivalent bi-partitions of
the system into two subsystems having, respectively, $m$ and $n-m$
qubits. The sum appearing in equation (\ref{Entsub}) has,
therefore, $K_{bipart}^{(m)}$ terms. $E({\rm bipart.})$ stands for
the entanglement associated with one single bi-partition of the
$n$-qubit system.

The above procedure for evaluating a global entanglement measure
on the basis of bi-partitions can be implemented in a variety of
ways, depending on which measure $E({\rm bipart.})$ one uses to
evaluate the entanglement of the individual bi-partitions. When
dealing with mixed states the negativity constitutes the most
important and most widely used, practical bipartite entanglement
measure. The negativity $N({\rm bipart.})$ corresponding a given
bipartition of the system into two subsystems, one with  $m$
qubits and the other one with $n-m$ qubits, with $m\le n-m$,
is given (up to a normalization multiplicative factor) by the sum
of the negative eigenvalues $\alpha_i$ of the partial transpose
matrix associated with this bipartition,

\begin{equation}\label{neg}
N({\rm bipart.})=
\frac{2}{2^m-1} \,
\sum_i |\alpha_i|.
\end{equation}

In this work we also consider an indicator of entanglement
$E_{MB}$ proposed by Mintert and Buchleitner \cite{MB07} which is
 given by
\begin{equation}
E_{MB}[\rho]=2Tr[\rho^2]-Tr[\rho_1^2]-Tr[\rho_2^2],
\end{equation}
with $\rho_1$ and $\rho_2$ the reduced density matrices
of the associated subsystems. The entanglement indicator $E_{MB}$
constitutes a lower bound of the squared concurrence, which for
mixed states $\rho$ is defined as \cite{MKB04}
\begin{equation}
C[\rho]=\inf\sum_ip_iC(\Psi_i)
\end{equation}
where $\{p_i,\Psi_i\}$ ($\sum_ip_i=1$) are all the
possible decompositions of $\rho$ into mixtures of pure states
$|\Psi_i\rangle$, and $C$ is the concurrence of a pure state of a
bipartite system,
\begin{equation}
C(\Psi)=\sqrt{2[1-Tr(\rho_r^2)]}
\end{equation}
\noindent whit $\rho_r$ the marginal density matrix corresponding
to either subsystem.

\subsection{Pure states}

In this section we determine numerically the decay of the average
entanglement of a multiqubit system when considering general,
random initial pure states uniformly generated according to the
Haar measure. All exhibited curves are obtained generating (and
averaging over) $10^4$ random pure states. See the Appendix A for
technical details on this choice of the sample size. The main
results obtained by us are summarized in figures 1-4. In figures
1a-1d we plot (for systems of three, four, five and six qubits
respectively) the average entanglement $\langle N\rangle$ and the
dispersion $\Delta N=\sqrt{\langle N^2\rangle-\langle N\rangle^2}$
as a function of the dimensionless quantity $\gamma _0t$.  The
dispersion $\Delta N$ gives an estimation of the width of the
entanglement distribution associated with initial random states.
In all cases it is seen that $\Delta N$ is relatively small
compared with $\langle N\rangle$, which means that this last
quantity is representative of the typical entanglement dynamics
exhibited by the multiqubit system. In figure 1 we also depict the
curves corresponding to the entanglement decay of some important
particular states, such as the $n$-qubit GHZ state (which is the
maximally entangled state in the three qubit case) and the
maximally entangled states for four, five, and six qubit systems
(the corresponding curves in figures 1c-1d are identified with the
label ``robust'', for reasons explained below). For systems of
four qubits we compute the entanglement evolution of the state HS,
which was introduced by Higuchi and Sudbery in \cite{HS00} (see
also \cite{BPBZCP07}). This state has been conjectured to maximize
the entanglement of four-qubit states \cite{HS00} and, even if a
proof of this conjecture is still lacking, various analytical and
numerical evidences supporting it have been reported in the
literature \cite{HS00,BSSB05,BPBZCP07}. In the cases of five and
six qubit-systems we plot the entanglement decay for the maximally
entanglement states $|\Phi^{ME}_{5qb}\rangle$ \cite{BSSB05} and
$|\Phi^{ME}_{6qb}\rangle$ \cite{BPBZCP07}. The HS four-qubit state
and the $|\Phi^{ME}_{6qb}\rangle$ six qubit states are specially
relevant in connection with the process of decoherence because
they also constitute robust states under decoherence. In a
previous work \cite{BMPCP08} we have numerically determined these
robust states for systems of four, five, and six qubits under
different (Markovian) decoherence processes. By a ``robust state"
we mean an initial state evolving under decoherence to states that
are, at any given time, more entangled than those generated by
alternative initial states.  The five qubit state
$|\Phi^{ME}_{5qb}\rangle$  does not coincide with the robust state
reported in \cite{BMPCP08}, but in the case of the dynamics
studied in the present work, our numerical results indicate that
both states lead to the same (or very close) average entanglement
evolution. \\

\begin{figure}
\begin{center}
\vspace{0.5cm}
\includegraphics[scale=0.8,angle=0]{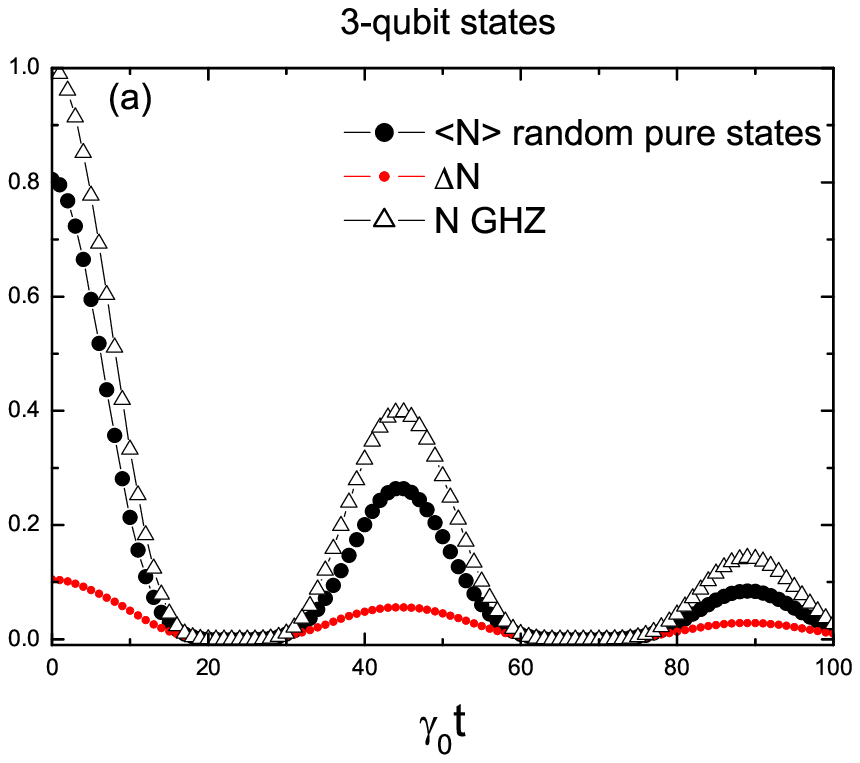}
\vspace{0.5cm}
\includegraphics[scale=0.8,angle=0]{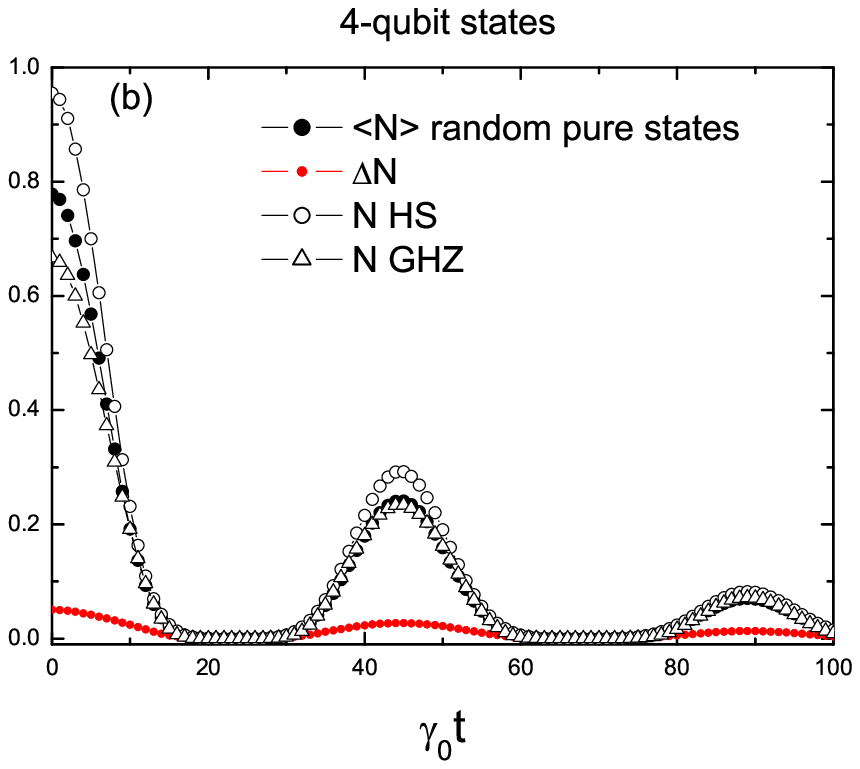}
\vspace{0.5cm}
\includegraphics[scale=0.8,angle=0]{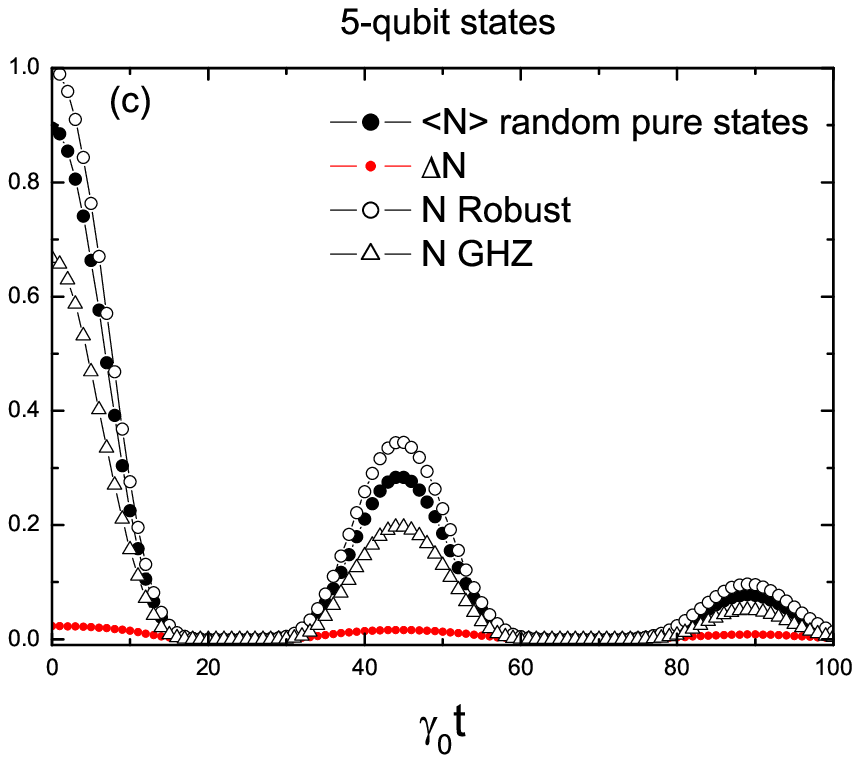}
\vspace{0.5cm}
\includegraphics[scale=0.8,angle=0]{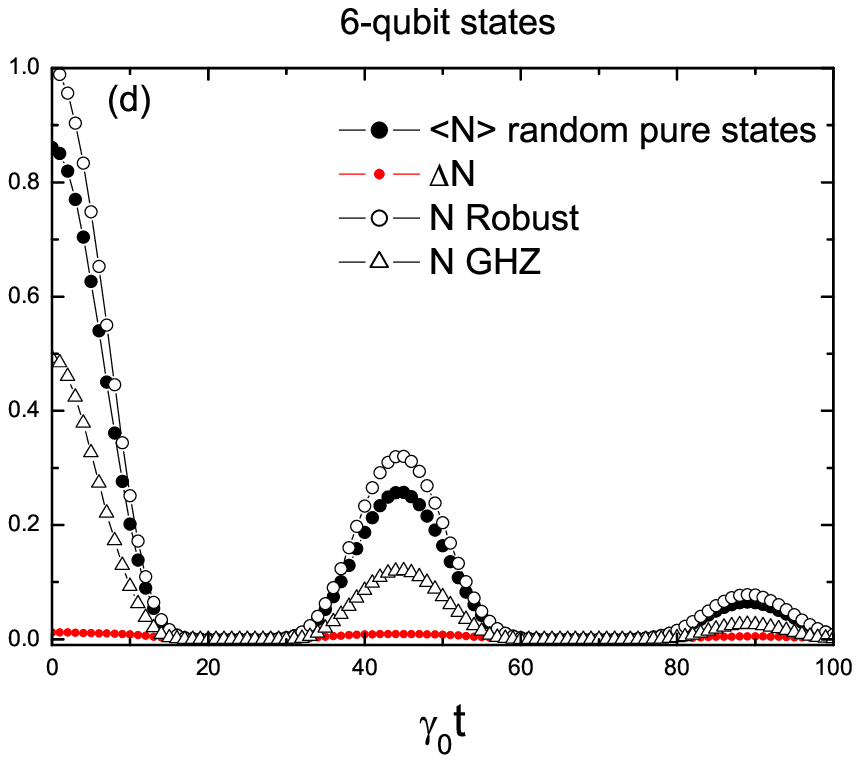}
\caption{Entanglement evolution of the 3, 4, 5 and 6 qubits random
(pure) states under decoherence. All depicted quantities are
dimensionless.\label{figu_1}}
\end{center}
\end{figure}

A particularly interesting aspect of figures 1a-1d is the status
of the state GHZ with respect to the typical behaviour of the
entanglement dynamics of the multiqubit system. In the case of
three-qubit systems the time dependent average entanglement is
appreciably below the entanglement of states that evolved from the
GHZ one, particularly at those times when the system's
entanglement reaches its maximum values. The situation is
different in the case of four-qubits, where we found that the
entanglement dynamics associated with a GHZ initial state is quite
representative of the typical, average dynamics associated with
random initial states. On the other hand, for systems of five and
six qubits the entanglement evolution associated with the GHZ
state does not reflect the average behaviour, which is more
closely represented by the entanglement dynamics exhibited by
states respectively evolving from the previously mentioned
$|\Phi^{ME}_{5qb}\rangle$ and  $|\Phi^{ME}_{6qb}\rangle$ states.
\\

\begin{figure}
\begin{center}
\vspace{0.5cm}
\includegraphics[scale=0.8,angle=0]{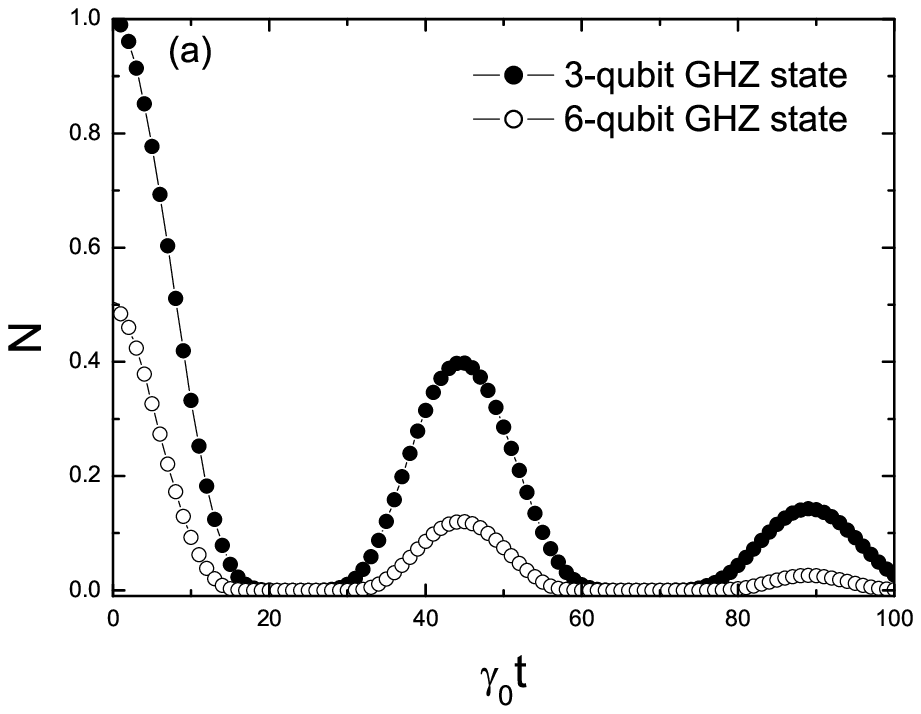}
\vspace{0.5cm}
\includegraphics[scale=0.8,angle=0]{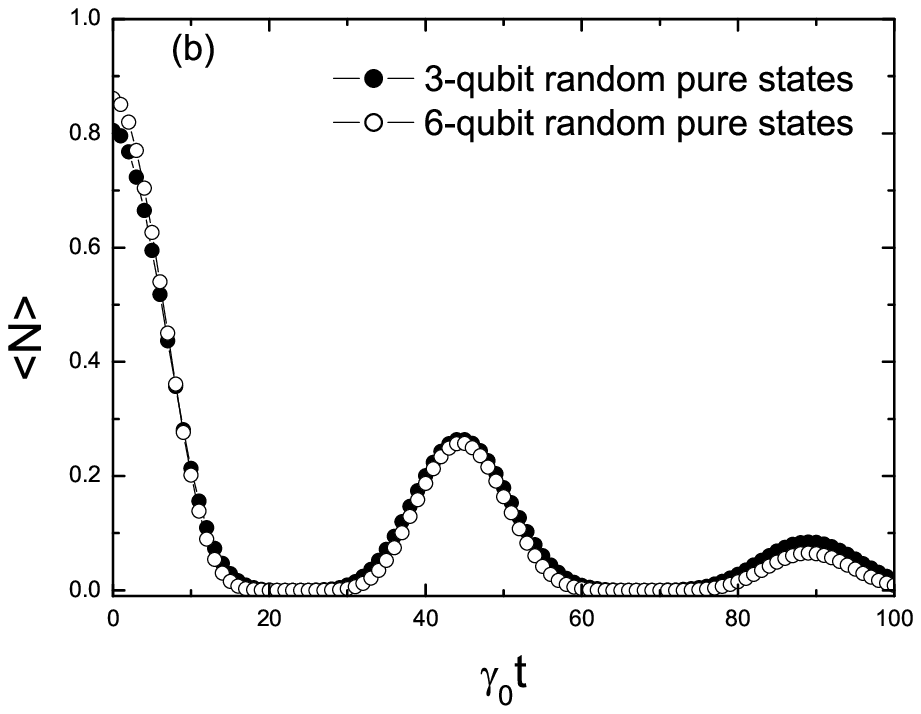}
\caption{Entanglement evolution of the 3-qubit and 6-qubit
representative states under decoherence: (a) GHZ states of three
and six qubits; (b) random pure states of three and six qubits.
All depicted quantities are
dimensionless.
\label{figu_2}}
\end{center}
\end{figure}

Evidence has been previously reported suggesting that in some
situations the robustness of entanglement may increase with the
number of qubits in the system \cite{SK02}. However, recent
investigations \cite{ACCAD08} indicate that, even when
entanglement takes a long time to disappear, it rapidly adopts
values too small to be of any practical relevance for
technological applications. It is interesting to explore the
dependence of the entanglement robustness on the number of qubits
within the context of the non-Markovian evolution studied in the
present work. Some results pertaining to this issue are summarized
in figures 2 and 3. In figure 2.a we compare the entanglement
evolution yielded by initial GHZ states of systems of three and
six qubits.  It transpires from this figure that the finite time
intervals of zero entanglement are longer in the case of six
qubits than in the case of three qubits. This suggests that the
lengths of these intervals tend to increase with the number of
qubits, consistently with other results for particular states
recently reported in \cite{HIPZ}. This trend, however, is not
observed when we evaluate the average entanglement evolution
corresponding to random initial pure states, as can be appreciated
in figure 2.b. Furthermore, the average entanglement evolutions
corresponding to random initial pure states of systems consisting
of 3, 4, 5, and 6 qubits are depicted in figure 3, showing an
almost universal pattern. The average entanglement of random
initial pure states thus seems to be less fragile than the
entanglement corresponding to the GHZ state. It is interesting to
compare these findings with those reported in a recent study by
Aolita et al. \cite{AolitaEtAl09}. These authors investigated the
decay of the entanglement under various decoherence channels of the
GHZ state. Aolita et al. determined an upper bound for the
entanglement of states evolved (under the alluded decoherence
channels) from initial GHZ states. They found, however, that the
average entanglement of states evolved from random initial pure
states does not comply with this bound, suggesting that the
exponentially increasing fragility of the GHZ entanglement with
the number of qubits of the system is not a generic feature of
typical pure states. Our present results constitute further
evidence also pointing towards the conclusion that the exponential
(in terms of system's size) fragility of the GHZ state may be a
peculiarity of this state not shared by generic ones. On the
contrary, general pure states, on average, seem to exhibit more
robust entanglement than the GHZ state. Another finding that is
relevant in connection with this problem was reported by Hein,
D\"ur and Briegel in \cite{HDB05}, where it was shown that the
entanglement decay under decoherence corresponding to certain graph states
states is very different from the one characterizing the GHZ states.
The entanglement decay associated with graph states with a constant degree
(not dependent on the number of qubits) does not become
more acute as the size of the system increases. Results indicating
that the entanglement fragility of the GHZ states is not a generic
feature of multi-qubit states were also presented by Carvalho,
Mintert and Buchleitner \cite{CMB04} who showed that the rates
of (exponential) decay of entanglement of the $W$ sates are
independent of the number of qubits of the system in the cases
of coupling to a zero temperature bath and of dephasing. On the
contrary, under these environmental influences the entanglement
decay rate grows linearly with the number of qubits for the GHZ state.

\begin{figure}
\begin{center}
\vspace{0.5cm}
\includegraphics[scale=0.95,angle=0]{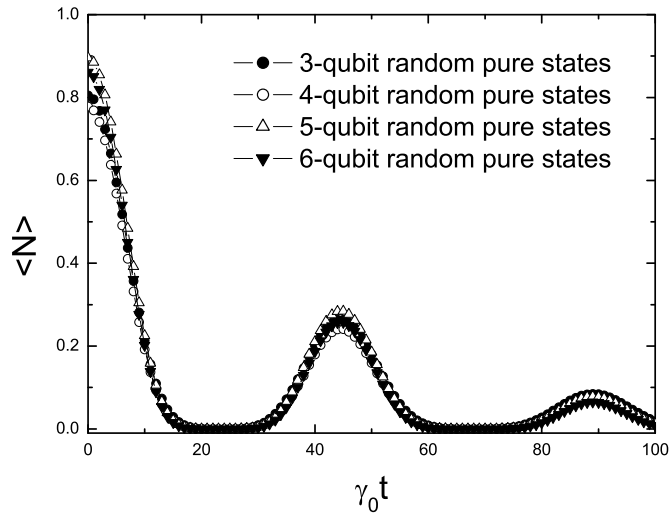}
\caption{Average entanglement evolution under decoherence
corresponding to random initial pure states of systems
of 3,4,5, and 6 qubits.
All depicted quantities are
dimensionless.
\label{figu3}}
\end{center}
\end{figure}

\begin{figure}
\begin{center}
\vspace{0.5cm}
\includegraphics[scale=0.95,angle=0]{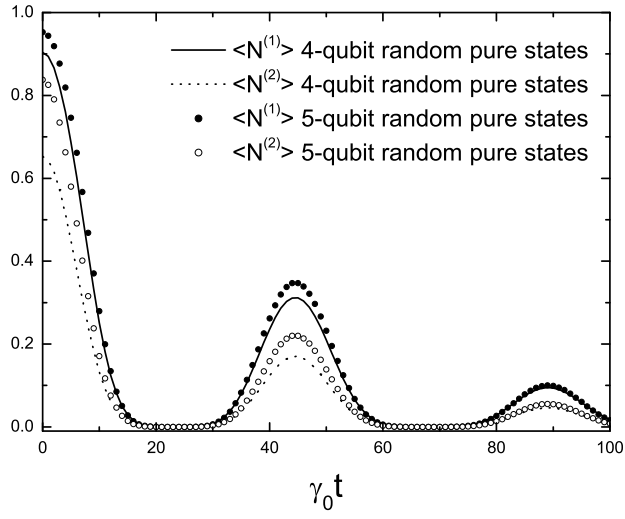}
\caption{Entanglement evolution of the random 4 (lines) and 5
(dots) qubit systems under decoherence for the most unbalanced
(solid line, full dots) and most balanced (dotted line, empty
dots) possible bipartitions of the systems. All depicted
quantities are dimensionless. \label{figu_3p}}
\end{center}
\end{figure}

\begin{figure}
\begin{center}
\vspace{0.5cm}
\includegraphics[scale=0.8,angle=0]{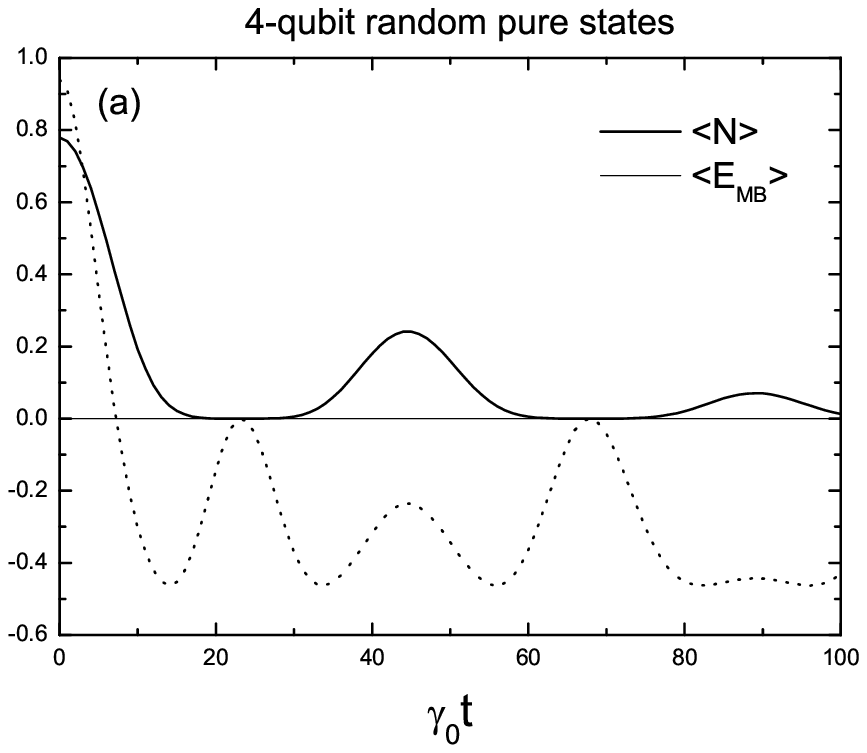}
\vspace{0.5cm}
\includegraphics[scale=0.8,angle=0]{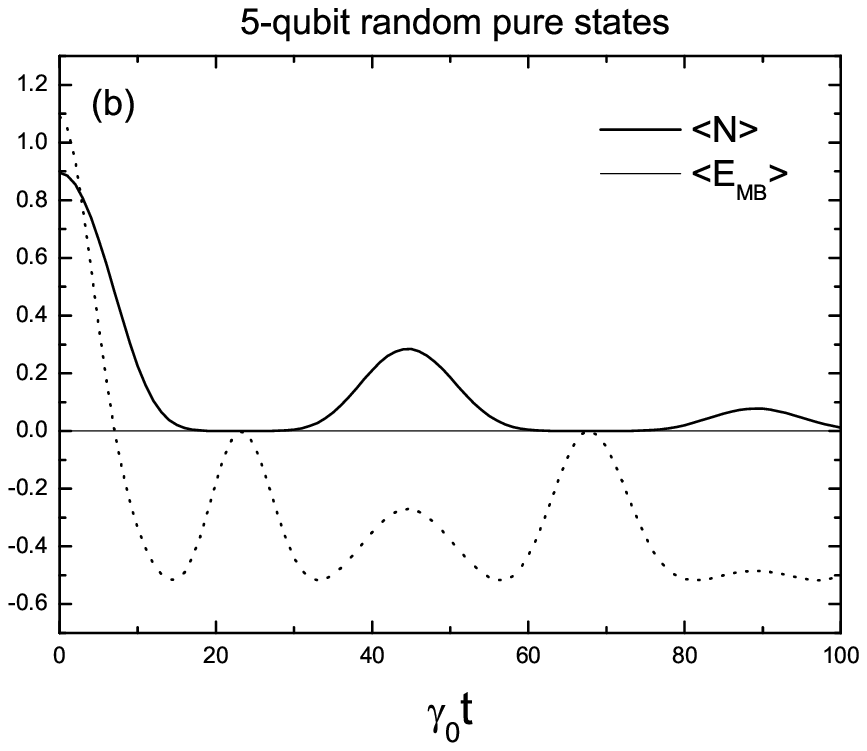}
\caption{The negativity and the $E_{MB}$ entanglement
indicator as a function of $\gamma_0 t$ for random 4
(a) and 5 (b) initially pure qubit systems. All depicted
quantities are dimensionless. \label{fig_3pp}}
\end{center}
\end{figure}

Let us denote by $\langle N^{(m)} \rangle$ the average negativity
corresponding to all the partitions of the system into two
subsystems having, respectively, $m$ and $n-m$ qubits. In figure
\ref{figu_3p} we compare for 4 and 5 qubit systems the evolution
of the entanglement of the most balanced bipartitions $\langle
N^{(2)}\rangle$ and the most unbalanced ones $\langle
N^{(1)}\rangle$. We can see that the behavior exhibited by the
negativities corresponding to these special bipartitions is very
similar to that of the global negativity $\langle N\rangle$. \\

The dynamical evolution of the average $\langle E_{MB}\rangle$ is
compared with that of the negativity in figure \ref{fig_3pp}, for 4
and 5 random qubit systems. In both cases the quantity $E_{MB}$ is
 able to detect entanglement only during the first period of
entanglement decrease, but it does not detect the subsequent
entanglement revivals. This is consistent with the fact that the
quality of the entanglement indicator $E_{MB}$ tends to
deteriorate as one considers states of increasing mixedness \cite{BMPCP09b},
and the multi-qubit systems under consideration here have a
considerable degree of mixedness during the entanglement revival
events. \\

It is known that the entanglement exhibited by composite quantum
systems tends to decrease as we consider states with increasing
degree of mixedness. It is interesting to notice that the average
entanglement evolution of the system under consideration in the
present work does not follow this general tendency. The average
time evolution of both the negativity and the degree of mixedness
(as measured by the linear entropy $S_L$ of the time dependent
state) corresponding to random initial pure states of six qubits
are depicted in figure 6. We can see that, except for the initial
entanglement decrease, the moments of high entanglement tend to
coincide with moments of high mixedness. This point was recently
highlighted in \cite{MMPS10}, in connection with the evolution of
systems of two-qubits, and is also consistent with previous
results advanced in \cite{HIPZ}. Before discussing in more detail
this aspect of the evolution of the multi-qubit systems discussed
here it is worth mentioning that this behavior is atypical but
should not be regarded as paradoxical. Even if, as a general
tendency, entanglement tends to decrease when the degree of
mixedeness increases, there is no strict correlation between these
two quantities. It is indeed possible to find pairs of states such
that the state with higher mixedness is also the more entangled
one. In spite of this, it is still interesting that the system's
evolution considered here consists (during certain time intervals)
of states showing this somewhat atypical characteristic.

\begin{figure}
\begin{center}
\vspace{0.3cm}
\includegraphics[scale=0.8,angle=0]{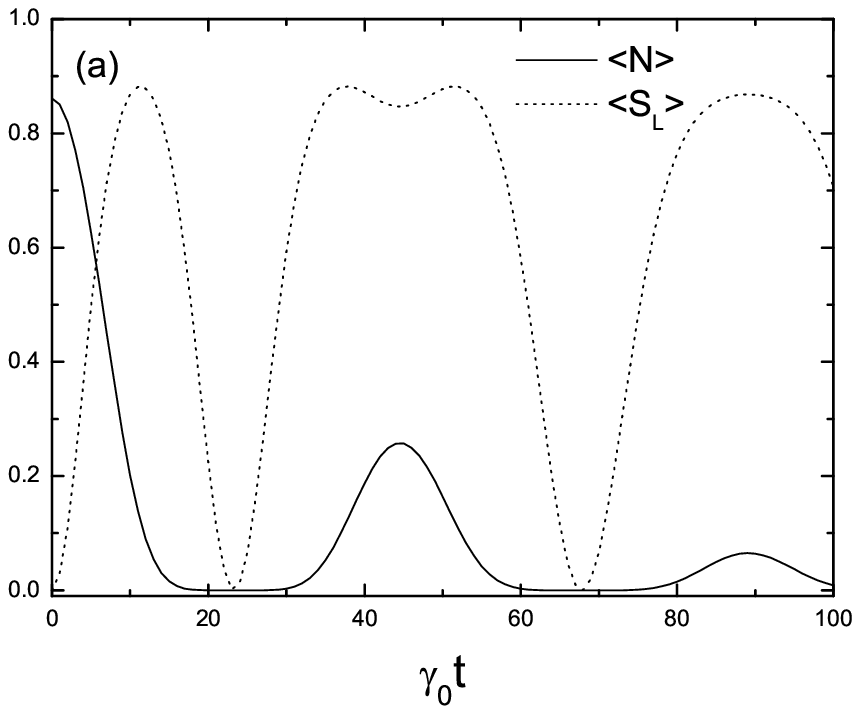}
\vspace{0.3cm}
\includegraphics[scale=0.8,angle=0]{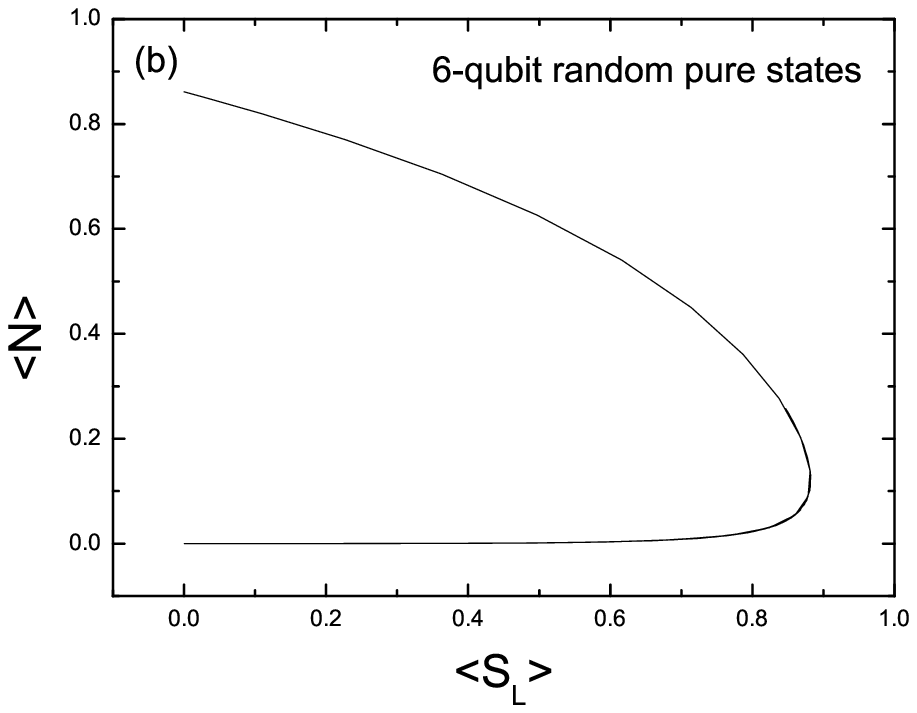}
\caption{Average time evolution of the global entanglement and the
linear entropy corresponding to random initial 6-qubit pure states
(a) and the mean global entanglement vs. the mean linear entropy
associated with those states (b). All depicted quantities are
dimensionless. \label{NvsSL}}
\end{center}
\end{figure}

To understand better this peculiarity let us first notice that the
system's entanglement vanishes at those times when the parameter
$p_t$ becomes equal to zero, $p_t=0$. At these times the mixedness
of the multi-qubit system also vanishes, because its state is
given by $\rho_s = (|0 \rangle\langle 0|)^{\otimes n}$, which is
pure and separable. When starting from an initial entangled pure
state $\rho_i = |\Psi\rangle\langle \Psi |$, the evolution of the
system's mixedness can be construed as arising from two opposite
tendencies. First, due to the interaction with its surroundings,
some entanglement is developed between the system and the
environment and the system's mixedness tends to increase. On the
other hand, as the evolution proceeds, the system eventually
begins to approach the pure state $\rho_s$, and its mixedness,
after having achieved a maximum value, starts to decrease again.
Consequently, the resulting evolution comprises an initial phase
of monotonous increase in the degree of mixedness followed by a
period of decreasing mixedness during which its value goes back to
zero. On the other hand, during this whole time interval before
$p_t$ achieves its first zero, the system's entanglement decreases
monotonously, finally vanishing when $p_t=0$. Notice that during
the second half of this period, during which the system is
evolving towards $\rho_s$, the system's entanglement and mixedness
both decrease together and finally vanish when $p_t=0$. After
this, when $p_t$ starts increasing again during the first
entanglement revival, the system's entanglement and degree of
mixedness grow together, reaching their respective maximum values
approximately at the same time and then decreasing together as the
system evolves back to the state $\rho_s$. \\

\subsection{Mixed Initial States}

Here we consider the entanglement decay of states of the form
(\ref{mixed-states}), consisting of a statistical mixture of a
pure state $|\Psi\rangle$ and the totally mixed state. Our results
are given in figure 7, where the average negativity $\langle N
\rangle$ is plotted against $\gamma_0 t$ and the parameter $x$ for
systems of three and six qubits. We see that the general patterns
associated with these two cases are similar except for the fact
that the average negativity of systems having six qubits decreases
more slowly with $x$ than the average negativity corresponding to
systems of three qubits.
\begin{figure}
\begin{center}
\vspace{0.5cm}
\includegraphics[scale=0.7,angle=0]{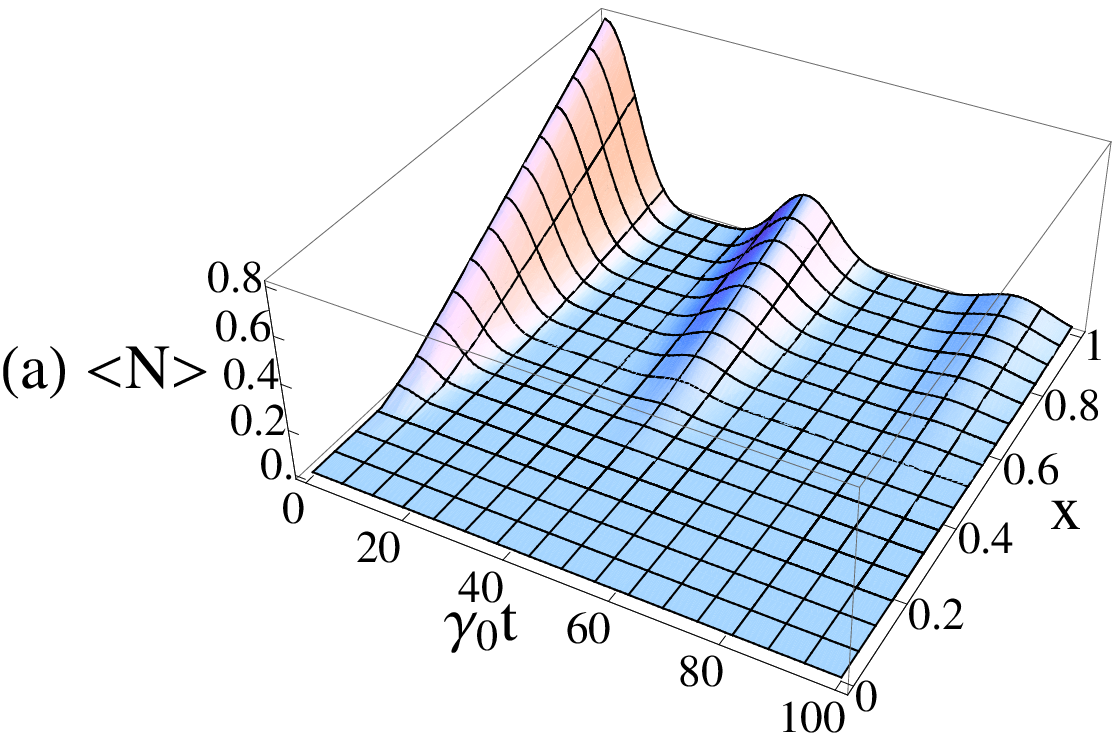}
\vspace{0.5cm}
\includegraphics[scale=0.7,angle=0]{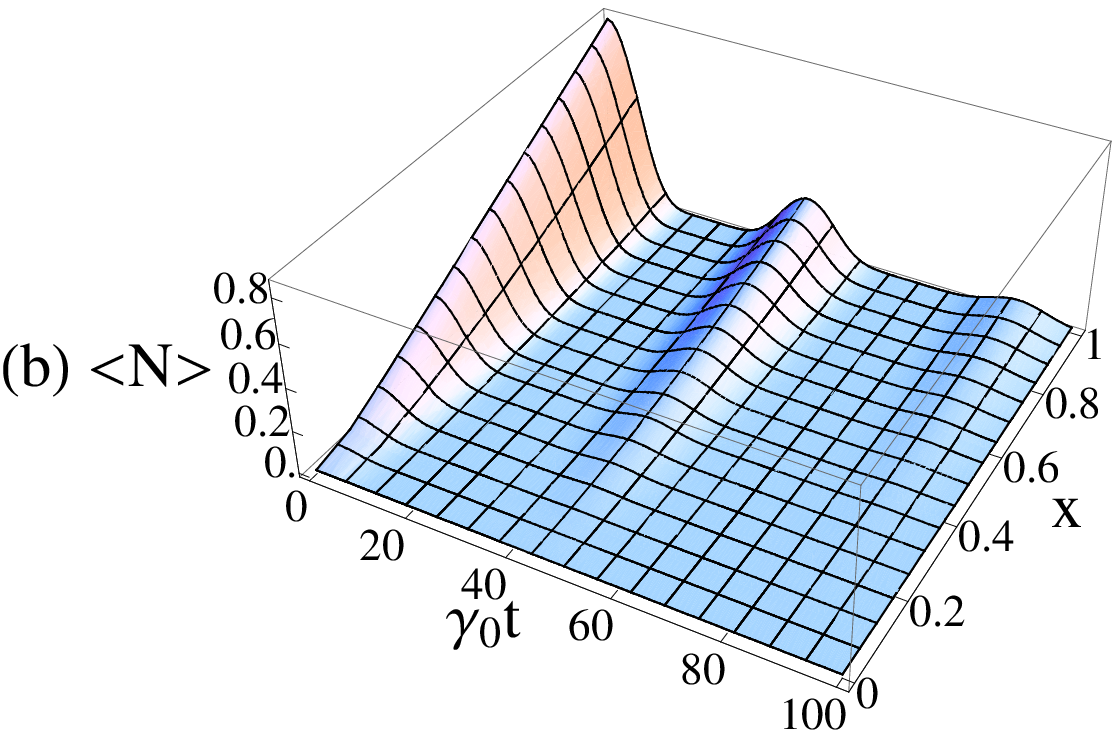}
\caption{Time evolution of $\langle N\rangle$ for random
initial $n$-qubit mixed states of the form (\ref{mixed-states})
as a function of $\gamma_0t$ and $x$, for $n=3$ (left) and $n=6$
(right).
All depicted quantities are
dimensionless.
\label{figu_4}}
\end{center}
\end{figure}
\section{Conclusions}

When considering the average behaviour of generic initial pure
states of multiqubit systems for $n=3, 4, 5,$ and $6$ qubits one
obtains in each case a very similar, almost universal pattern of
entanglement evolution. This is an interesting result, specially
in connection with some previous studies suggesting that
entanglement becomes more robust as the size of the system
increases. It is clear that the system under study here does not
follow that trend. Nor does the average entanglement become
appreciably more fragile as the number of qubits increases, at
least within the range of $n$-values considered by us.\\

In the cases of systems having $5$ and $6$ qubits, the robust
states studied in \cite{BMPCP08} exhibit, under the dynamics
considered in the present work, a behaviour that is quite close to
the average behaviour corresponding to generic initial pure
states. In this sense, the entanglement dynamics of the alluded
robust states is representative of the typical dynamics of initial
pure states. On the contrary, the average entanglement dynamics
of $5$ and $6$ qubits departs from the one exhibited by the
corresponding GHZ states, whose entanglement appears to be more
fragile than the entanglement of typical, average, pure states.
These findings are consistent with other recent results in the
literature, also indicating that the entanglement fragility of
GHZ states under various decoherence channels is not a generic
feature of typical multi-qubit pure states. \\

Systems of four qubits show some peculiarities. In the case of
these systems the average entanglement evolution of random initial
pure states is quite similar to the entanglement evolution corresponding
to an initial GHZ state. \\

The observed universal features exhibited by the average entanglement
dynamics associated with random initial pure states of $n$-qubits is
plausibly related to the fact that each qubits is interacting with its
``own'' independent heat bath. This fact implies that the typical behaviours
of systems with different (total) numbers $n_1$ and $n_2$ of qubits,
even though they are not identical, share
some important features: the evolution rules for the local states of
 blocks of $L$ qubits (with $L\le \min(n_1,n_2)$) are the same in both cases.
For instance, the evolution equation for single qubits is always given by
equation (\ref{dinunqu}). \\

The fact that the GHZ states deviate from the typical behaviour, having
 entanglement features that are more fragile than those of typical
initial states, is consistent with other known manifestations of
the ``fragility'' of the quantum correlations exhibited by these
states. For instance, it is well-known that if we trace over one
of the qubits of a system initially described by a GHZ state, the
remaining $n-1$ qubits are left in a mixed non-entangled state.
Also, if we measure the state of one of the qubits, the remaining
qubits are left in a non-entangled state. These properties are in
contrast with the ones exhibited by other multi-qubit entangled
states, such as the W states, where some entanglement survives
after measuring the state of one of the parts of the system. As a
final remark concerning GHZ states, it is worth mentioning that
for systems having five or more qubits the difference between the
entanglement of the maximally entangled states (according to the
partitions-based measures considered here) and the GHZ states is
much larger than in the case of systems of four qubits. This may
be related to the fact that the behaviour of GHZ states deviate
more from the average behaviour in the case of systems with five
or more qubits, than in
the case of systems with four qubits. \\

The main results presented in this work have been obtained using
entanglement measures based upon the negativities associated with
the different possible bi-partitions of the multi-qubit system. We
focused mostly on the behavior exhibited by the global mean value
corresponding to all those bi-partitions. However, the
entanglement dynamics corresponding to particular bi-partitions
was also considered, and the results obtained are similar to those
corresponding to the global average. The negativity, even though
it provides one particular way of quantitatively characterizing
the amount of entanglement in composite quantum systems, it
constitutes one of the most widely used {\it practical} tools for
studying the entanglement of mixed states. It would be interesting
to explore other ways of characterizing the evolution of the
entanglement (or other manifestations of quantum mechanical
correlations) of the system considered here. Any further
developments along these
lines will be very wellcome.\\


\section*{Acknowledgements}
This work was partially supported by the Projects FQM-2445 and
FQM-207 of the Junta de Andalucia, and the grant FIS2011-24540 of
the Ministerio de Innovaci\'on y Ciencia (Spain). A.P.M.
acknowledges support by GENIL trhough YTR-GENIL Program.

\appendix
\section{Details of the Numerical Calculations}

In this appendix we provide some technical details on the
numerical Monte Carlo procedure employed to compute the average
properties corresponding to general initial pure states of the
multi-qubit system. As was stated in the main text of this work,
we compute these averages by generating a random sample of $10^4$ initial
pure sates. These states are generated according to the Haar
measure (see \cite{BCPP02} and references therein). The Haar measure
determines a uniform distribution in the space of $2^n\times 2^n$
unitary matrices which, in turn, induces a natural uniform distribution
on the Hilbert $2^n$-dimensional space corresponding to the $n$-qubit
system. To generate uniformly distributed random initial states $|\Psi\rangle$
we generate random $2^n\times 2^n$ unitary matrices $ U^{(n)}$
uniformly distributed according to the Haar measure,
and then compute $|\Psi\rangle \, = \, U^{(n)} |\Psi_0 \rangle$,
where $|\Psi_0 \rangle$ is a fixed $n$-qubit state.

To ensure that we have an appropriate sample, we consider
increasing values of its size until the computed averages do not change
any more (within a certain range of numerical error) with further increases
of the sample size. The behavior of the numerically determined
average negativity $\langle N \rangle$ as a function of the sample
size, corresponding to the case of four qubits and $\gamma_0
t=40$, is depicted as an illustration in figure 8. The behavior of
$\langle N \rangle$ against the sample size corresponding to
systems of three, five and six qubits is plotted in the inset of
figure 8. It transpires from figure 8 that the changes on the
numerical values of $\langle N \rangle$ corresponding to an
increase of the sample size from $10^4$ to $10^5$ are already
negligible within the scale of the figures exhibited in the main
text of this paper.

\begin{figure}
\begin{center}
\vspace{0.5cm}
\includegraphics[scale=0.9,angle=0]{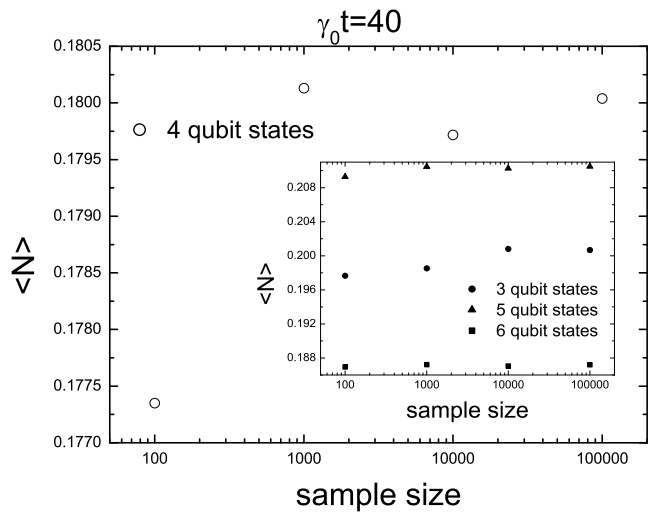}{B}
\caption{$\langle N\rangle$ as a function of the sample size for 4
qubit random states. Inset: $\langle N\rangle$ as a function of
the sample size for 3 (circle), 5 (triangle), and 6 (square) qubit
random states. All depicted quantities are dimensionless.
\label{fig_8}}
\end{center}
\end{figure}

\end{document}